\documentclass[notitlepage,prd,showpacs,superscriptaddress,nofootinbib,floatfix,showkeys,10pt]{revtex4-1}
\usepackage{graphicx}
\usepackage{amsmath}
\usepackage{bm}
\usepackage{yhmath}
\usepackage{mathtools}
\usepackage{wasysym}
\usepackage{subfigure}
\usepackage{xcolor}
\definecolor{CherryRed}{rgb}{.65,0,.2}
\definecolor{RubyRed}{rgb}{.88,0.07,.3}
\definecolor{CralRed}{rgb}{1,0.25,.25}
\definecolor{CobaltBlue}{rgb}{0,0.28,.67}
\definecolor{RoyalBlue}{rgb}{0.25,0.41,.88}
\definecolor{EmeraldGreen}{rgb}{0.31,0.78,.47}
\definecolor{EmeraldGreen}{rgb}{0.31,0.78,.47}
\definecolor{LimeGreen}{rgb}{50,205,50}
\definecolor{ForestGreen}{RGB}{34,139,34}
\definecolor{PineGreen}{RGB}{1,121,111}\usepackage{cases}
\usepackage[colorlinks=true, linkcolor=EmeraldGreen,citecolor=RoyalBlue,urlcolor=ForestGreen,linkcolor=ForestGreen]{hyperref}
\usepackage{subfigure}
\usepackage{times}
\usepackage{booktabs,bm}
\usepackage{slashed}
\usepackage{amsfonts,amssymb,stmaryrd,latexsym,amsmath}
\usepackage{textcomp}
\usepackage{multirow}
\usepackage{cancel}
\usepackage{array}
\usepackage{hyperref}
\usepackage{pgf}
\usepackage{orcidlink}

\def\SU3{{\text{SU(3)}_{\rm F}}}

\def \pcs4338{{P_{\psi s}^\Lambda(4338)^0}}

\begin{document}
	
	\title{\textcolor{CobaltBlue}{  Strong vertex $N  N \rho$  in dense medium}}

	\author{M.~Shekari Tousi\orcidlink{0009-0007-7195-0838}}
	\email{ marzie.sh.tousi@ut.ac.ir }
	\affiliation{Department of Physics, University of Tehran, North Karegar Avenue, Tehran 14395-547,  Iran }

	\author{K.~ Azizi\orcidlink{0000-0003-3741-2167}}
	\email{kazem.azizi@ut.ac.ir}
	\thanks{Corresponding author}
	\affiliation{Department of Physics, University of Tehran, North Karegar Avenue, Tehran 14395-547,  Iran }
	\affiliation{Department of Physics, Faculty of Engineering and Natural Sciences, Dogus University, Dudullu-\"{U}mraniye, 34775 Istanbul, T\"urkiye }

	\date{\today}

	\begin{abstract}

We investigate the strong  coupling form factors associated with the $NN\rho$ interaction vertex in nuclear matter using the QCD sum rule framework. These form  factors are calculated incorporating the in-medium modifications of the relevant condensates and operators into the operator product expansion (OPE) for the first time. We analyze the behavior of the strong form factors as functions of the momentum transfer squared and density of medium and evaluate the corresponding strong couplings at the on-shell point of the $\rho$ meson at nuclear saturation density. Then we extend the analyses to get information at higher densities.
 Our results provide complementary information on the in-medium $NN\rho$ interaction and may be useful for phenomenological, experimental and observational studies of dense hadronic matter.

\end{abstract}

	\maketitle
	\renewcommand{\thefootnote}{\#\arabic{footnote}}
	\setcounter{footnote}{0}

\section {Introduction}\label{sec:one}
Quantum Chromodynamics (QCD) provides a fundamental framework for describing the strong interaction and the internal structure of hadrons. While the perturbative regime is well-understood, the nonperturbative domain remains a central challenge, particularly when addressing the properties of hadrons in a nuclear matter environment. Understanding how strong-interaction vertices, which govern the dynamics of nuclear forces, are modified at finite density is crucial for a microscopic description of dense nuclear matter and the interior of neutron stars.

The properties of hadrons in a nuclear medium have been a subject of extensive research. Specifically, the in-medium modifications of nucleon properties \cite{Cohen:1991nk,Furnstahl:1992pi,Azizi:2014yea}, as well as those of light vector mesons like the $\rho$ meson \cite{Hatsuda:1991ez,Asakawa:1993wh,Klingl:1997kf}, have been systematically investigated through various nonperturbative approaches, reflecting their critical role in nuclear dynamics. However, these studies primarily focus on the individual characteristics of  the hadrons, such as their in-medium masses, decay constants, and spectral distributions. Despite the extensive literature on strong vertices in the vacuum \cite{Azizi:2014bua,Azizi:2015tya,Azizi:2015bxa,Azizi:2015jya,Agaev:2019qqn,Aliev:2009ei,Olamaei:2020bvw,Azizi:2020zin,Azizi:2021utt},
 there is a notable lack of studies concerning the density dependence of the interaction vertices governing meson-baryon dynamics in a medium.
 In particular, the $NN\rho$ interaction is an important  vertex in nuclear matter; nevertheless, the   strong coupling constant ($g_{NN\rho}$)  remains largely unexplored from a fundamental QCD perspective in a dense nuclear medium. 
In the vacuum, the strong coupling constants  associated with the $NN\rho$ interaction vertex have been extensively investigated   using  QCD sum rules \cite{Aliev:2009ei,Wang:2007yt,Zhu:1998yga,Erkol:2006sa}. These vacuum investigations have provided valuable insight into the strong coupling constant by analyzing three-point correlation functions or employing the light-cone QCD sum rule technique. 

In effective descriptions of nuclear interactions, the isovector-vector $\rho$ meson plays an important role in the isospin-dependent part of the nucleon-nucleon interaction. In relativistic mean-field (RMF) models, the $\rho$ meson-nucleon coupling contributes to the description of the nuclear symmetry energy and its density dependence, which are relevant to the properties of asymmetric nuclear matter and neutron stars \cite{Machleidt:1987hj,Serot:1984ey,Horowitz:2000xj,Steiner:2004fi,Typel:2009sy}. The in-medium modification of the $NN\rho$ coupling constant  is therefore of interest for understanding how the effective nucleon-meson interaction changes in dense nuclear matter. In many effective nuclear models, the density-dependent meson-nucleon couplings are introduced through phenomenological parameterizations or determined by fitting nuclear matter properties \cite{Typel:1999yq,Lalazissis:2005de}.

The QCD sum rule framework is a powerful, nonperturbative approach that connects hadronic observables to the underlying QCD parameters \cite{Shifman:1978bx,Shifman:1978by}. By employing the operator product expansion  and matching correlation functions with hadronic representations, this method provides a systematic way to incorporate the density dependence of QCD vacuum condensates. While the QCD sum rule approach has been successfully applied to characterize the in-medium modifications of individual nucleons and mesons, its application to the strong coupling constants of meson-baryon interaction vertices in nuclear matter remains largely unexplored.

Indeed, the calculation of the $g_{NN\rho}$ coupling constant within a finite density environment represents a significant gap in the current literature. Most existing studies are either restricted to vacuum conditions or rely on phenomenological models that lack a direct connection to the underlying QCD dynamics. Consequently, a first-principles analysis, incorporating the in-medium modifications of the relevant condensates, is essential to establish a rigorous theoretical foundation for the $NN\rho$ vertex in dense nuclear matter.

In the present work, we address this gap by investigating the $NN\rho$ vertex using an in-medium three-point correlation function. By constructing a suitable correlation function and accounting for the density-dependent modifications of the condensates, we obtain a quantitative determination of how this fundamental vertex is modified by the surrounding nuclear environment. Our results offer   theoretical insights and serve as a vital input for future investigations into hadronic interactions in nuclear matter.

The rest of this article is structured as follows: The formalism, grounded in the QCD sum rule framework and the associated computational configuration, is detailed in Section \ref{sec:two}. We proceed in Section \ref{sec:three} to derive the strong coupling constant $g_{NN\rho}$ for the $NN\rho$ vertex, subsequently analyzing our results. Finally, Section \ref{sec:four} is dedicated to summarizing our findings and evaluating the broader physical implications of this work.

\section {The strong coupling constants of $NN\rho$   in nuclear matter via QCD sum rules }\label{sec:two}

This section outlines the theoretical framework adopted to determine the strong coupling constants between the nucleon states and the $\rho$ meson via the QCD sum rule approach in nuclear matter. Our formulation is built upon the following three-point in-medium correlation function composed of the interpolating currents associated with the interacting hadrons:
\begin{eqnarray}\label{CorrelationFunction}
	\Pi_{\mu}(p,q)=i^2 \int d^4x~ \int d^4y~e^{-ip\cdot x}~
	e^{ip^{\prime}\cdot y}~{\langle}0| {\cal T}\left (
	J^{N_2}(y)~J_{\mu}^{\rho}(0)~\bar{J}^{N_1}(x)\right)|0{\rangle}_{\rho_N},
\end{eqnarray}
where $\rho_N$ is the nuclear matter density, ${\cal T}$ stands for time ordering, and $q = p - p'$ denotes the transferred momentum four-vector. In this setup, $J^i$ denotes the corresponding interpolating current of the nucleon or the $\rho$ meson. The labels $N_1$ and $N_2$  denote the initial and final nucleon states.  The analysis of this correlator proceeds through two distinct regimes: the phenomenological (hadronic) representation and the short-distance operator product expansion. Equating the coefficients of identical Lorentz structures from both sides allows us to isolate the desired strong coupling parameters. Finally, to eliminate contaminating effects from higher excited resonances and the continuum spectrum, we apply a double Borel transformation with respect to the virtualities $-p^2$ and $-p'^2$, along with the implementation of continuum threshold subtractions.

\subsection{Hadronic Side}

On the hadronic representation of the correlation function, the interpolating currents act as creation and annihilation operators for the corresponding hadronic states. To carry out the phenomenological evaluation, complete sets of intermediate hadronic states possessing quantum numbers identical to those of the interpolating currents are inserted into the correlator. Performing the four-dimensional spacetime integrations leads to the following expression:
\begin{eqnarray}
	\Pi _{\mu }^{\mathrm{Had}}(p, q)=\frac{\langle 0|J^{N_2 }|N_2(p',s')\rangle \langle 0|J_{\mu }^{\rho}|\rho(q)\rangle \langle \rho(q) N_2(p',s')|N_1(p,s)\rangle \langle N_1(p,s)|J^{N_1}|0\rangle _{\rho_N}}{(p'^2-m_{N_2}^{*2})(q^2-m_{\rho}^{*2})(p^2-m_{N_1}^{*2})}+\cdots,  \label{eq:CorrF1Phys}
\end{eqnarray}
where the ellipsis ($\cdots$) parameterizes the contributions originating from higher excitations and continuum states and $m_{N}^{*}$ and $m_{\rho}^{*}$ are the in-medium mass of the nucleon and $\rho$ meson. The transition matrix elements entering this expression are conventionally parameterized in terms of the in-medium masses, residues and decay constant  as follows:
\begin{eqnarray}
	\langle 0|J^{N_1}|N_1(p,s)\rangle _{\rho_N}&=&\lambda_{N_1}^* u_{N_1}(p,s),
	\nonumber\\
	\langle 0|J^{N_2 }|N_2(p',s')\rangle _{\rho_N}&=&\lambda_{N_2}^* u_{N_2}(p',s'),
	\nonumber\\
	\langle 0|J_{\mu }^{\rho}|\rho(q)\rangle _{\rho_N}&=&f_{\rho}^*  m_{\rho}^* \varepsilon_{\mu},
	\label{eq:ResPc}
\end{eqnarray}
in which $\varepsilon_{\mu}$ denotes the polarization vector, and $f_{\rho}^*$ represents the  decay constant of the $\rho$ meson in nuclear matter. Here, $\lambda_{N_1}^*$ and $\lambda_{N_2}^*$ correspond to the in-medium residues associated with the $N_1$ and $N_2$ states, respectively, with $u_{N_1}$ and $u_{N_2}$ being their corresponding Dirac spinors. The single-nucleon state is denoted by $|N_i(p,s)\rangle$. Furthermore, the  vertex matrix element can be parameterized in terms of the coupling constants $g_1$ and $g_2$ as \cite{Azizi:2021utt}:
\begin{eqnarray}
	\langle \rho(q) N_2(p',s')|N_1(p,s)\rangle _{\rho_N} = \varepsilon^{* \nu}\bar{u}_{N_2}(p',s')\big[g_1\gamma_{\nu}-\frac{i\sigma_{\nu\alpha}}{m_{N_2}^*+m_{N_1}^*}q^{\alpha}g_2\big] u_{N_1}(p,s).
	\label{eq:coupling}
\end{eqnarray}
Subsequently, substituting the matrix elements defined in Eqs.~(\ref{eq:ResPc}) and (\ref{eq:coupling}) into Eq.~(\ref{eq:CorrF1Phys}) and performing spin summations over the Dirac spinors and polarization vectors via the standard completeness relations:
\begin{eqnarray}
	\sum_{s}u_{N_1}(p,s)\bar{u}_{N_1}(p,s)&=&({\not\!p}+m_{N_1}^*),\nonumber \\
	\sum_{s'}u_{N_2}(p',s')\bar{u}_{N_2}(p',s')&=&({\not\!p'}+m_{N_2}^*), \nonumber\\
	\sum_{\lambda}\varepsilon^{\lambda}_{\alpha}\varepsilon^{\lambda  *}_{\beta}&=&-g_{\alpha\beta}+\frac{q_\alpha q_\beta}{m_{\rho}^{*2}},
	\label{eq:SumPc}
\end{eqnarray}
 we eventually arrive at an expression that consists of the product of the following form
\begin{equation}
	\Pi_{\mu}^{Had}(p,q)=\frac{(\!\not\!{p'}+m_{N_2}^{*}) \cdots (\!\not\!{p}+m_{N_1}^{*})}{(p'^2-m_{N_2}^{*2})(p^2-m_{N_1}^{*2})}+ \cdots =\frac{1 }{(\!\not\!{p'}-m_{N_2}^{*})} \cdots \frac{1 }{(\!\not\!{p}-m_{N_1}^{*})}+\cdots.
\end{equation}
To account for the interaction between the nucleon and the nuclear medium, the hadronic representation of the correlation function is modified to incorporate the in-medium self-energies and we express as
\begin{equation}\label{corre1}
\frac{1 }{(\!\not\!{p}-m_{N_1}^{*})}=\frac{1}{
		(p^\mu-\Sigma_\nu^\mu)\gamma_\mu-m_{N_1}^*  }+\cdots,
\end{equation}
and 
\begin{equation}\label{corre2}
	\frac{1 }{(\!\not\!{p'}-m_{N_2}^{*})}=\frac{1}{
		(p'^\mu-\Sigma_\nu^\mu)\gamma_\mu-m_{N_2}^* }+\cdots,
\end{equation}
where $\Sigma_\nu^\mu$ denotes the vector   self-energy of the nucleon in the presence of nuclear matter \cite{Cohen:1991js}. In general, the vector self-energy is decomposed as
\begin{equation}\label{sigma}
	\Sigma_\nu^\mu=\Sigma_\nu u^\mu+\Sigma'_\nu p^\mu,
\end{equation}
where $\Sigma_\nu$ and $\Sigma'_\nu$ are constants and $u^\mu$ represents the four-velocity of the nuclear medium. Consistent with previous studies \cite{Cohen:1991js}, we neglect the $\Sigma'_\nu$ term given its negligible contribution. Unlike vacuum QCD sum rule calculations, the introduction of the medium's four-velocity provides a distinct frame of reference that necessitates the emergence of additional structures in the correlation function. By operating in the rest frame of the nuclear matter, where  $u^\mu = (1, \mathbf{0})$, and substituting Eq. (\ref{sigma}) into Eq. (\ref{corre1}), the hadronic side of the correlation function is rewritten as the product of the following form \cite{Cohen:1994wm}
\begin{equation}
	\Pi_{\mu}^{Had}(p,q)=\frac{1}{
		(\!\not\!{p'}-\Sigma_\nu\!\not\!{u})-m_{N_2}^* } \cdots\frac{1}{
		(\!\not\!{p}-\Sigma_\nu\!\not\!{u})-m_{N_1}^* }+\cdots.
\end{equation} 
This expression can be expanded in terms of some independent Lorentz structures. Finally, we implement the double Borel transformation with respect to the momentum variables to exponentially suppress the continuum and higher resonance contributions. Consequently, the hadronic side of the correlation function takes the following form:

\begin{eqnarray}
	\tilde{\Pi} _{\mu }^{\mathrm{Had}}(Q^2)&=&e^{-\frac{{\mu_1}^2}{M^2}}e^{-\frac{{\mu_2}^2}{M'^2}}\frac{f_{\rho}^* m_{\rho}^* \lambda_{N_2}^* \lambda_{N_1}^*  }{2(m_{\rho}^{*2} + Q^2)}\big[ g_2   ({\mu_1}^2 - {\mu_2}^2-2 Q^2) 
+ 2 g_1 (m_{N_1}^* m_{N_2}^*- {\mu_2}^2- Q^2+ 2 p_0 \Sigma_\nu - \Sigma_{\nu}^2) \big]\gamma_{\mu} \nonumber\\ 
	&-&e^{-\frac{{\mu_1}^2}{M^2}}e^{-\frac{{\mu_2}^2}{M'^2}}\frac{f_{\rho}^* m_{\rho}^* \lambda_{N_2}^* \lambda_{N_1}^*  }{2 m_{N_1}^* (m_{\rho}^{*2} + Q^2)}\big[2  g_1    m_{N_1}^*  m_{N_2}^*  
	+  g_2 (m_{N_1}^* m_{N_2}^*+ {\mu_1}^2-  2 p_0 \Sigma_\nu + \Sigma_{\nu}^2) \big] \not\!q \gamma_{\mu}  \nonumber\\
	&+&\mathrm{other~structures}+\cdots ,
	\label{eq:had}
\end{eqnarray}
where $M^2$ and $M'^2$ are the Borel mass parameters introduced by the transformation, which are to be fixed by fulfilling standard sum rule criteria, and $Q^2 = -q^2$ is the space-like momentum transfer squared. Here, ${\mu_1}^2=m_{N_1}^{*2}-\Sigma_{\nu}^2+2p_0\Sigma_\nu$ and ${\mu_2}^2=m_{N_2}^{*2}-\Sigma_{\nu}^2+2p'_0\Sigma_\nu$, where $p_0$  and $p'_0$  correspond to the quasi-particle energies. 
Although the physical dispersion relation accommodates additional independent  structures beyond those explicitly shown in Eq.~(\ref{eq:had}), we retain solely the structures directly employed in our numerical analysis, whereas the remaining structures  are subsumed under $\mathrm{other~structures}+\cdots$.

\subsection{QCD Side}
The theoretical evaluation of the correlation function, within the deep Euclidean regime, relies on the Operator Product Expansion. We define the interpolating field for the nucleon as:
\begin{eqnarray}\label{InterpolatingCurrents}
	J_{N}(y)=2\varepsilon_{ij\ell}\Big\{\Big(u^{i^T}(y)Cd^{j}(y)\Big)\gamma_{5}
	u^{\ell}(y)+\beta\Big(u^{i^T}(y)C\gamma_5d^{j}(y)\Big)
	u^{\ell}(y)\Big\},
\end{eqnarray}
which facilitates the coupling to the nucleon states. The corresponding current for the $\rho$ meson is constructed as:
\begin{eqnarray}
J_{\mu}^{\rho}(0)=\frac{1}{\sqrt{2}}\Big(\bar{u}(0)\gamma_{\mu}u(0)-\bar{d}(0)\gamma_{\mu} d(0)\Big).
\end{eqnarray}
Within Eq.~(\ref{InterpolatingCurrents}), $C$ denotes the charge conjugation matrix, while $\beta$ serves as a generalized mixing parameter; for instance, the specific choice of $\beta=-1$ recovers the standard Ioffe current. The value of $\beta$ is subsequently determined by invoking relevant phenomenological constraints. Substituting these explicit operator forms into the correlator defined in Eq.~(\ref{CorrelationFunction}) necessitates the contraction of all quark fields through the application of Wick’s theorem, an operation that yields:
\begin{eqnarray}\label{correlfuncOPE}
	\Pi_{\mu}^{QCD}(p,q)&=&2i^2\int d^{4}x\int d^{4}ye^{-ip\cdot
		x}e^{ip^{\prime}\cdot y}\epsilon_{abc}\epsilon_{ij\ell}
	\nonumber \\
	&\times&
	\Bigg\{-Tr\Big[S_d^{bj}(y-x)S_u^{\prime^{di}}(-x)\gamma_{\mu}S_u^{\prime^{ad}}(y)\Big]\Big(\gamma_5S_u^{c\ell}(y-x)\gamma_5\Big)
+ Tr\Big[S_u^{\prime^{ai}}(y-x)S_d^{bj}(y-x)\Big]\Big(\gamma_5S_u^{cd}(y)\gamma_{\mu}S_u^{d\ell}(-x)\gamma_5\Big)
	\nonumber \\
	&-&\gamma_5S_u^{ci}(y-x)S_d^{\prime^{bj}}(y-x)S_u^{ad}(y)\gamma_{\mu}S_u^{d\ell}(-x)\gamma_5
-\gamma_5
	S_u^{cd}(y)\gamma_{\mu}S_u^{di}(-x)S_d^{\prime^{bj}}(y-x)S_u^{a\ell}(y-x)\gamma_5
	\nonumber \\
	&-&Tr\Big[S_u^{\prime^{ai}}(y-x)S_d^{bd}(y)\gamma_{\mu}S_d^{dj}(-x)\Big]\Big(\gamma_5S_u^{c\ell}(y-x)\gamma_5\Big)	-\gamma_5S_u^{ci}(y-x)S_d^{\prime^{dj}}(-x)\gamma_{\mu}S_d^{\prime^{bd}}(y)S_u^{a\ell}(y-x)\gamma_5
	\nonumber \\
	&-&\beta\Bigg[Tr\Big[S_d^{bj}(y-x)\gamma_{\mu}S_u^{\prime^{di}}(-x)\gamma_5S_u^{\prime^{ad}}(y)\Big]\Big(\gamma_5S_u^{c\ell}(y-x)\Big) 
	+Tr\Big[\gamma_5S_d^{bj}(y-x)S_u^{\prime^{di}}(-x)\gamma_{\mu}S_u^{\prime^{ad}}(y)\Big]\Big(S_u^{c\ell}(y-x)\gamma_5\Big)
	\nonumber \\
	&-&Tr\Big[\gamma_5S_u^{\prime^{ai}}(y-x)S_d^{bj}(y-x)\Big]\Big(\gamma_5S_u^{cd}(y)\gamma_{\mu}S_u^{d\ell}(-x)\Big)
-Tr\Big[S_u^{\prime^{ai}}(y-x)\gamma_5S_d^{bj}(y-x)\Big]\Big(S_u^{cd}(y)\gamma_{\mu}S_u^{d\ell}(-x)\gamma_5\Big)
	\nonumber \\
	&+&\gamma_5S_u^{ci}(y-x)\gamma_5S_d^{\prime^{bj}}(y-x)S_u^{ad}(y)\gamma_{\mu}S_u^{d\ell}(-x)
+S_u^{ci}(y-x)S_d^{\prime^{bj}}(y-x)\gamma_5S_u^{ad}(y)\gamma_{\mu}S_u^{d\ell}(-x)\gamma_5
	\nonumber \\
	&+&S_u^{cd}(y)\gamma_{\mu}S_u^{di}(-x)S_d^{\prime^{bj}}(y-x)\gamma_5S_u^{a\ell}(y-x)\gamma_5
+\gamma_5S_u^{cd}(y)\gamma_{\mu}S_u^{di}(-x)\gamma_5S_d^{\prime^{bj}}(y-x)S_u^{a\ell}(y-x)
	\nonumber \\
	&+&Tr\Big[\gamma_5S_u^{\prime^{ai}}(y-x)S_d^{bd}(y)\gamma_{\mu}S_d^{dj}(-x)\Big]\Big(\gamma_5S_u^{c\ell}(y-x)\Big)
+Tr\Big[S_u^{\prime^{ai}}(y-x)\gamma_5S_d^{bd}(y)\gamma_{\mu}S_d^{dj}(-x)\Big]\Big(S_u^{c\ell}(y-x)\gamma_5\Big)
	\nonumber \\
	&+&S_u^{ci}(y-x)S_d^{\prime^{dj}}(-x)\gamma_{\mu}S_d^{\prime^{bd}}(y)\gamma_5S_u^{a\ell}(y-x)\gamma_5
+\gamma_5S_u^{ci}(y-x)\gamma_5S_d^{\prime^{dj}}(-x)\gamma_{\mu}S_d^{\prime^{bd}}(y)S_u^{a\ell}(y-x)
	\Bigg]
	\nonumber \\
	&-&\beta^2\Bigg[Tr\Big[\gamma_5S_d^{bj}(y-x)\gamma_5S_u^{\prime^{di}}(-x)\gamma_{\mu}S_u^{\prime^{ad}}(y)\Big]S_u^{c\ell}(y-x)
-Tr\Big[\gamma_5S_u^{\prime^{ai}}(y-x)\gamma_5S_d^{bj}(y-x)\Big]\Big(S_u^{cd}(y)\gamma_{\mu}S_u^{d\ell}(-x)\Big)
	\nonumber \\
	&+&S_u^{ci}(y-x)\gamma_5S_d^{\prime^{bj}}(y-x)\gamma_5S_u^{ad}(y)\gamma_{\mu}S_u^{d\ell}(-x)
+S_u^{cd}(y)\gamma_{\mu}S_u^{di}(-x)\gamma_5S_d^{\prime^{bj}}(y-x)\gamma_5S_u^{a\ell}(y-x)
	\nonumber \\
	&+&Tr\Big[\gamma_5S_u^{\prime^{ai}}(y-x)\gamma_5S_d^{bd}(y)\gamma_{\mu}S_d^{dj}(-x)\Big]S_u^{c\ell}(y-x)
+S_u^{ci}(y-x)\gamma_5S_d^{\prime^{dj}}(-x)\gamma_{\mu}S_d^{\prime^{bd}}(y)\gamma_5S_u^{a\ell}(y-x)
	\Bigg]
	\Bigg\}~,
\end{eqnarray}
where $S^{\prime}=CS^TC$. Here, $S^{ij}_{q}(x)$
corresponds to the light quark propagator in nuclear matter for which we utilized
\cite{Cohen:1994wm,Reinders:1984sr}
\begin{eqnarray}\label{propagator}
	S_{q}^{ab}(x)&\equiv& \langle0|  {\cal T} [q^a
	(x)\bar{q}^b(0)]|0\rangle_{\rho_N} \nonumber \\
	&=&
	\frac{i}{2\pi^2}\delta^{ab}\frac{1}{(x^2)^2}\not\!x
	-\frac{m_q }{ 4\pi^2} \delta^ { ab } \frac { 1}{x^2} \nonumber \\
	&+&
	\chi^a_q(x)\bar{\chi}^b_q(0)-\frac{ig_s}{32\pi^2}F_{\mu\nu}^A(0)t^{ab,A
	}\frac{1}{x^2}[\not\!x\sigma^{\mu\nu}+\sigma^{\mu\nu}\not\!x]+\cdots,
\end{eqnarray}
where $q$ denotes $u$ or $d$ quark and $\chi^a_q$ and $\bar{\chi}^b_q$ are the Grassmann-valued background quark fields, whereas $F_{\mu\nu}^A$ represents the classical external gluon field. The first two terms in the expansion of the light quark propagator account for the perturbative contributions, including the first-order correction in the quark mass. Conversely, the subsequent two terms encompass the non-perturbative dynamics stemming from the background quark and gluon condensates, respectively. Specifically, the gluonic contribution arises from a single-gluon interaction, where only the leading-order term in the short-distance operator product expansion is retained. Following standard procedure (see, e.g., Ref. \cite{Cohen:1994wm}), we neglect contributions arising from derivatives of the gluon field strength tensor, as well as higher-order gluon interactions. Upon substituting the propagator expression of Eq.~(\ref{propagator}) into the correlation function given by Eq.~(\ref{CorrelationFunction}), we obtain products of the Grassmann background quark fields and the classical gluon fields, which are subsequently identified with the vacuum-to-ground-state matrix elements of the respective quark and gluon operators \cite{Cohen:1994wm}
\begin{eqnarray}\label{}
	&&\chi_{a\alpha}^{q}(x)\bar{\chi}_{b\beta}^{q}(0)=\langle
	q_{a\alpha}(x)\bar{q}_{ b\beta}(0)\rangle_{\rho_N}, ~~~~~~~
	F_{\kappa\lambda}^{A}F_{\mu\nu}^{B}=\langle
	G_{\kappa\lambda}^{A}G_{\mu\nu}^{B}\rangle_{\rho_N}, \nonumber \\
	&&\chi_{a\alpha}^{q}\bar{\chi}_{b\beta}^{q}F_{\mu\nu}^{A}=\langle
	q_{a\alpha}\bar{q}_{ b\beta}G_{\mu\nu}^{A}\rangle_{\rho_N}, 
	~~~~~~~~\chi_{a\alpha}^{q}\bar{\chi}_{b\beta}^{q}\chi_{c\gamma}^{q}\bar
	{\chi}_{d\delta}^{q}=\langle
	q_{a\alpha}\bar{q}_{b\beta}
	q_{c\gamma}\bar{q}_{d\delta}\rangle_{\rho_N}.
\end{eqnarray}
Before proceeding to the numerical analysis, it is necessary to formally define the quark, gluon, and mixed condensates within the nuclear medium. The operator product expansion depends on the in-medium matrix element of the quark propagator, which is projected as \cite{Cohen:1994wm}:
\begin{eqnarray} \label{ }
	\langle
	q_{a\alpha}(x)\bar{q}_{b\beta}(0)\rangle_{\rho_N}&=&-\frac{\delta_{ab}}{12}\Bigg
	[\Bigg(\langle\bar{q}q\rangle_{\rho_N}+x^{\mu}\langle\bar{q}D_{\mu}q\rangle_{
		\rho_N}
	+\frac{1}{2}x^{\mu}x^{\nu}\langle\bar{q}D_{\mu}D_{\nu}q\rangle_{\rho_N}
	+\cdots \Bigg)\delta_{\alpha\beta}\nonumber \\
	&&+\Bigg(\langle\bar{q}\gamma_{\lambda}q\rangle_{\rho_N}+x^{\mu}\langle\bar{q}
	\gamma_{\lambda}D_{\mu} q\rangle_{\rho_N}
	+\frac{1}{2}x^{\mu}x^{\nu}\langle\bar{q}\gamma_{\lambda}D_{\mu}D_{\nu}
	q\rangle_{\rho_N}
	+\cdots\Bigg)\gamma^{\lambda}_{\alpha\beta} \Bigg].\nonumber \\
\end{eqnarray}
Furthermore, the mixed quark-gluon condensate in the nuclear environment is expressed as:
\begin{eqnarray} \label{ }
	\langle
	g_{s}q_{a\alpha}\bar{q}_{b\beta}G_{\mu\nu}^{A}\rangle_{\rho_N}&=&-\frac{t_{
			ab}^{A
	}}{96}\Bigg\{\langle g_{s}\bar{q}\sigma\cdot
	Gq\rangle_{\rho_N}\Bigg[\sigma_{\mu\nu}+i(u_{\mu}\gamma_{\nu}-u_{\nu}\gamma_{\mu
	})
	\!\not\! {u}\Bigg]_{\alpha\beta} \nonumber \\
	&&+\langle g_{s}\bar{q}\!\not\! {u}\sigma\cdot
	Gq\rangle_{\rho_N}\Bigg[\sigma_{\mu\nu}\!\not\!
	{u}+i(u_{\mu}\gamma_{\nu}-u_{\nu}\gamma_{\mu}
	)\Bigg]_{\alpha\beta} \nonumber \\
	&&-4\Bigg(\langle\bar{q}u\cdot D u\cdot D q\rangle_{\rho_N}+im_{q}\langle\bar{q}
	\!\not\! {u}u\cdot D q\rangle_{\rho_N}\Bigg) \nonumber \\
	&&\times\Bigg[\sigma_{\mu\nu}+2i(u_{\mu}\gamma_{\nu}-u_{\nu}\gamma_{\mu}
	)\!\not\! {u}\Bigg]_{\alpha\beta}\Bigg\},
\end{eqnarray}
where $t_{ab}^{A}$ denotes the Gell-Mann matrices and $D_\mu=\frac{1}{2}(\gamma_\mu \!\not\!{D}+\!\not\!{D}\gamma_\mu)$. Additionally, the four-dimension gluon condensate matrix element is parameterized by:
\begin{equation}
	\langle
	G_{\kappa\lambda}^{A}G_{\mu\nu}^{B}\rangle_{\rho_N}=\frac{\delta^{AB}}{96}
	\Bigg[
	\langle
	G^{2}\rangle_{\rho_N}(g_{\kappa\mu}g_{\lambda\nu}-g_{\kappa\nu}g_{\lambda\mu}
	)+O(\langle
	\textbf{E}^{2}+\textbf{B}^{2}\rangle_{\rho_N})\Bigg],
\end{equation}
in which the final term is neglected due to its negligible contribution. The relevant in-medium condensates are defined according to \cite{Cohen:1994wm,Jin:1992id}:
\begin{eqnarray} \label{ }
	\langle\bar{q}\gamma_{\mu}q\rangle_{\rho_N}&=&\langle\bar{q}\!\not\!{u}q\rangle_
	{\rho_N}
	u_{\mu} , \\
	\langle\bar{q}D_{\mu}q\rangle_{\rho_N}&=&\langle\bar{q}u\cdot D
	q\rangle_{\rho_N}
	u_{\mu}=-im_{q}\langle\bar{q}\!\not\!{u}q\rangle_{\rho_N}
	u_{\mu}  ,\\
	\langle\bar{q}\gamma_{\mu}D_{\nu}q\rangle_{\rho_N}&=&\frac{4}{3}\langle\bar{q}
	\!\not\! {u}u\cdot D q\rangle_{\rho_N}(u_{\mu}u_{\nu}-\frac{1}{4}g_{\mu\nu}) 
	+\frac{i}{3}m_{q} \langle\bar{q}q\rangle_{\rho_N}(u_{\mu}u_{\nu}-g_{\mu\nu}),
	\\
	\langle\bar{q}D_{\mu}D_{\nu}q\rangle_{\rho_N}&=&\frac{4}{3}\langle\bar{q}
	u\cdot D u\cdot D q\rangle_{\rho_N}(u_{\mu}u_{\nu}-\frac{1}{4}g_{\mu\nu}) 
	-\frac{1}{6} \langle
	g_{s}\bar{q}\sigma\cdot Gq\rangle_{\rho_N}(u_{\mu}u_{\nu}-g_{\mu\nu}) ,\\
	\langle\bar{q}\gamma_{\lambda}D_{\mu}D_{\nu}q\rangle_{\rho_N}&=&2\langle\bar{q}
	\!\not\! {u}u\cdot D u\cdot D
	q\rangle_{\rho_N}\Bigg[u_{\lambda}u_{\mu}u_{\nu} -\frac{1}{6} 
	(u_{\lambda}g_{\mu\nu}+u_{\mu}g_{\lambda\nu}+u_{\nu}g_{\lambda\mu})\Bigg]\nonumber\\
	&&-\frac{1}{6} \langle
	g_{s}\bar{q}\!\not\! {u}\sigma\cdot
	Gq\rangle_{\rho_N}(u_{\lambda}u_{\mu}u_{\nu}-u_{\lambda}g_{\mu\nu}),
\end{eqnarray}
where the equations of motion are invoked, and terms of order $\mathcal{O}(m^2_q)$ are omitted owing to their infinitesimal impact \cite{Cohen:1994wm}. Having established the expressions for the light quark propagator and the various condensates, we substitute these into Eq. (\ref{correlfuncOPE}) and integrate over the coordinate space   to obtain the correlator in momentum space. 
To accomplish this transformation, we invoke the following identity, which is valid in $D$ spacetime dimensions:

\begin{eqnarray}\label{intyx}
	\frac{1}{[A^2]^n}&=&\int\frac{d^Dt}{(2\pi)^D}e^{-it\cdot A}~i~(-1)^{n+1}~2^{D-2n}~\pi^{D/2}~
	\frac{\Gamma(D/2-n)}{\Gamma(n)}\Big(-\frac{1}{t^2}\Big)^{D/2-n}.
\end{eqnarray}
Following the insertion of the propagator, the four-fold integrations over $x$ and $y$ are carried out by means of the replacements $x_{\mu}\rightarrow i\frac{\partial}{\partial p_{\mu}}$ and $y_{\mu}\rightarrow -i\frac{\partial}{\partial p'_{\mu}}$. Subsequently, upon employing the Feynman parametrization together with the master integral

\begin{eqnarray}\label{Int}
	\int d^4t\frac{(t^2)^{\beta}}{(t^2+L)^{\alpha}}=\frac{i \pi^2
		(-1)^{\beta-\alpha}\Gamma(\beta+2)\Gamma(\alpha-\beta-2)}{\Gamma(2)
		\Gamma(\alpha)[-L]^{\alpha-\beta-2}},
\end{eqnarray}
we arrive at the compact decomposition

\begin{eqnarray}\label{correlfuncOPELast}
\Pi_{\mu}^{QCD}(p,q)&=&\Pi_{p_{\mu} \not\!p}(p,q)p_{\mu} \not\!p+\Pi_{\not\!p \gamma_{\mu}}(p,q)\not\!p\gamma_{\mu}+\Pi_{\not\!q\gamma_{\mu}}(p,q)\not\!q\gamma_{\mu}
	+\Pi_{\gamma_{\mu}}(p,q) \gamma_{\mu}+\cdots,
\end{eqnarray}
in which the coefficient functions $\Pi_i(p,q)$ receive additive contributions from both the perturbative and non-perturbative sectors of the calculation. In terms of the double dispersion integrals, these functions admit the representation

\begin{eqnarray}\label{QCDside1}
	\Pi_i(p,q)=\int^{s_0}_{s_{min}}ds\int^{s'_0}_{s'_{min}}ds^{\prime}
	\frac{\rho_i^{pert}(s,s^{\prime},q^2)+\rho_i^{non-pert}(s,s^{\prime},q^2)}{(s-p^2)
		(s^{\prime}-p^{\prime^2})}~,
\end{eqnarray}
where $\rho_i(s,s',q^2)$ denote the spectral densities affiliated with the distinct Lorentz structures that emerge above. They are extracted by taking the imaginary part of the corresponding $\Pi_i$ functions, namely $\rho_i(s,s',q^2)=\frac{1}{\pi}\text{Im}[\Pi_{i}]$.   $s_0$ and $s'_0$ represent the continuum threshold. By way of illustration, we confine our attention here to the densities attached to  $\gamma_{\mu}$ and $\not\!q \gamma_{\mu} $ structures, which are found to read

\begin{eqnarray}\label{rho1pert}
	\rho_{\gamma_{\mu}}^{pert}(s,s^{\prime},q^2)&=& \int_{0}^{1}dj \int_{0}^{1-j}dv
	\frac{1}{128 \sqrt{2} j^3 \pi^4}\Bigg[\beta^2 \Bigg\{2 A^2 s' v^2 (m_d^2 + 6 m_u^2) + 
	2 j^3 v \Big(6 (5 q^2 - s - 5 s') m_d m_u 	\nonumber \\
	& + &B (7 D v^2 - 3 m_u^2)\Big) + 
	j v \Big(-11 L^2 + 16 A L s' v + 
	2 A (q^2 - s + s') v (m_d^2 + 6 m_u^2)\Big) - 
	2 j^2 \Bigg(-6 m_d m_u (C v + 6 m_u^2) 	\nonumber \\
	&+& m_d^2 (-q^2 v^2 + 6 m_u^2) + 
	v \Big[v \Big(8 L (s - s') - 7 A B D v\Big) - 
	3 [s - s' + q^2 (1 + 2 v)] m_u^2\Big]\Bigg)\Bigg\} + 
	2 \Bigg[2 A^2 s' v^2 (m_d^2 + 3 m_u^2)	\nonumber \\
	& +& 
	j^3 v \Big(8 B D v^2 + 3 (5 q^2 + s - 5 s') m_d m_u - 3 C m_u^2\Big) + 
	2 j v \Bigg(-4 L^2 + 5 A L s' v + 
	A v \Big((q^2 - s + s') m_d^2 - 3 s' m_d m_u	\nonumber \\
	& +& 
	3 (q^2 - s + s') m_u^2\Big)\Bigg) + 
	j^2 \Bigg(2 q^2 v^2 m_d^2 + 
	3 m_d m_u (v (-3 s + s' + 2 s v - q^2 (1 + 2 v)) + 4 m_u^2) + 
	v \Big(2 v [5 L (-s + s') 	\nonumber \\
	&+ &4 A B D v]+ 
	3 [s - s' + q^2 (1 + 2 v)] m_u^2\Big)\Bigg)\Bigg] - 
	2 \beta\Bigg( A^2 s' v^2 (m_d^2 - 7 m_u^2) + 
	j^3 v \Big(-2 B D v^2 + 6 D m_d m_u	\nonumber \\
	& + &3 (-5 q^2 + s + 5 s') m_u^2\Big) + 
	j^2 \Big(-3 m_d m_u (v (s - (3 q^2 + s) v) + 6 m_u^2) + 
	m_d^2 (q^2 v^2 + 30 m_u^2) + 
	v \Big[v [3 L (s - s') 	\nonumber \\
	&-& 2 A B D v] + [q^2 (3 - 4 v) + 
	s (6 + 9 v) - 3 (s' + 4 s' v)] m_u^2\Big]\Big) + 
	j v \Big[3 L^2 - 3 A L s' v + 
	A v \Big(s' (m_d^2 + 9 m_d m_u 	\nonumber \\
	& - &16 m_u^2)+ (q^2 - s) (m_d^2 - 
	7 m_u^2)\Big) \Big]\Bigg)  \Bigg] \Theta\Big[L(s,s^{\prime},q^2)\Big]
	,
\end{eqnarray}

\begin{eqnarray}\label{rho1nonpert}
	\rho_{\gamma_{\mu}}^{non-pert}(s,s^{\prime},q^2)&=& \int_{0}^{1}dj \int_{0}^{1-j}dv \Bigg[
	\frac{1}{24 \sqrt{2} j \pi^2}\Big[6 \Big(10 + \beta (3 + 8 \beta)\Big) j p_0 \langle q^{\dag}
	q\rangle_{\rho_N} v + 
	6 \langle\bar{q}q\rangle_{\rho_N} \Big(9 \beta j + \Big[-3 + \beta (9 + \beta)\Big] v\Big) m_d\nonumber \\
	& - &
	2 \Big(28 + \beta (15 + 14 \beta)\Big)\langle\bar{q}q\rangle_{\rho_N} m_q + 
	3 \langle\bar{q}q\rangle_{\rho_N} \Big((-4 + \beta + 2 \beta^2) j + 8 v + (9 - 10 \beta) \beta v \Big) m_u\Big] 	\nonumber \\
	&+&\frac{\langle
		g_{s}^{2}G^{2}\rangle_{\rho_N}}{12288 \sqrt{2} j^3 \pi^{4}} \Big[   \Big(81 + 3 (13 - 97 \beta) \beta \Big) j^2 - 
		4 (17 + 6 \beta + 8 \beta^2) (-1 + v) v - 
		j \Big(-39 + 83 v \nonumber \\
		& + &\beta [6 + 30 v + \beta (-33 + 71 v)]\Big) \Big]  \Bigg]  \Theta\Big[L(s,s^{\prime},q^2)\Big],
\end{eqnarray}

\begin{eqnarray}\label{rho2pert}
	\rho_{\not\!q \gamma_{\mu}}^{pert}(s,s^{\prime},q^2)&=& \int_{0}^{1}dj \int_{0}^{1-j}dv
	\frac{1}{64 \sqrt{2} j^2 \pi^4}\Bigg[v \Bigg(6 (-1 + \beta) (A + j) m_d^2 m_u + 
	m_u  \Bigg\{L \Big(-6 (-1 + 4 j + 3 v) + 
	\beta \Big[1 - 13 j \nonumber \\
	&+ &5 v + 6 \beta (-1 + 3 j + 3 v)\Big]\Big) + 
	3 v \Big[-2 A B v - 
	j \Big(s' (3 - 3 j - 5 v) + s (3 + j - 3 v) + 
	q^2 (-3 + 3 j + 5 v)\Big)\nonumber \\
	& + &
	2 \beta^2 \Big(A B v + 
	j \Big[C j + s + s' - (s + 2 s') v + q^2 (-1 + 2 v)\Big]\Big) + 
	\beta \Big(A B v + 
	j \{q^2 (3 - 3 j - 2 v) \nonumber \\
	&+ &s' (-3 + 3 j + 2 v) + 
	s (-3 + 4 j + 3 v)\}\Big)\Big] + 9 \beta (A + j) m_u^2  \Bigg\} + 
	m_d \Big[3 j (2 L + A B v + C j v) \nonumber \\
	&+ &
	\beta^2 \Big(3 L (1 + j - v) - A^2 B v + 
	j v \{2 C j + A q^2 + 3 s + s' - (2 s + s') v\}\Big) + 
	\beta \Big(L (13 - 25 j - 11 v) \nonumber \\
	&+ &
	v [3 s - 3 j \{(-1 + 3 j) q^2 + 2 (1 + j) s + s' - 3 j s'\} - 
	4 s v + j (-q^2 + 5 s + s') v + s v^2 \nonumber \\
	&- &
	A (q^2 - s') (-3 + 9 j + v)]\Big) + 
	3 \beta \Big(3 A + (15 + 14 \beta) j \Big) m_u^2 \Big] \Bigg)  \Bigg] \Theta\Big[L(s,s^{\prime},q^2)\Big]
	,
\end{eqnarray}

\begin{eqnarray}\label{rho2nonpert}
	\rho_{\not\!q \gamma_{\mu}}^{non-pert}(s,s^{\prime},q^2)&=& \int_{0}^{1}dj \int_{0}^{1-j}dv 
	\frac{6 + \beta (-3 + 14 \beta)}{8 \sqrt{2} \pi^2}\Big(  \langle\bar{q}q\rangle_{\rho_N} v \Big)  \Theta\Big[L(s,s^{\prime},q^2)\Big] ,
\end{eqnarray}
where the argument of the step function $\Theta$ and other parameters are specified through
\begin{eqnarray}\label{teta1}
	L(s,s^{\prime},q^2)&=&-\Big((-1 + j) j s\Big) -B  j   (-1 + j + v) + 
	s' (j - j^2 + v - 2 j v - v^2),
	\nonumber \\
	A&=&(-1 + v),
	\nonumber \\
B&=&q^2 - s - s',	\nonumber \\
C&=&q^2 + s - s',	\nonumber \\
D&=&q^2 - s'.
\end{eqnarray}

To exponentially suppress contributions from higher resonances and the continuum, we apply a double Borel transformation, supplemented by the quark-hadron duality assumption.  The derivation is then concluded by equating the hadronic and OPE representations, whereby the coefficients of identical Lorentz structures are matched term by term, so as to yield the desired QCD sum rules. This procedure yields a system of two coupled equations involving both $g_1$ and $g_2$. By solving these relations simultaneously and considering  $Q^2 = -q^2$, we analytically extract the $g_1$ and $g_2$ within the sum rule framework for    $\gamma_{\mu}$ and $ \not\!q \gamma_{\mu}$ structures, resulting in:

\begin{eqnarray}
	g_1(\rho_N,Q^2)&=&\frac{ -e^{\frac{{\mu_1}^2}{M^2}}e^{\frac{{\mu_2}^2}{M'^2}} (m_{\rho}^{*2} + Q^2) \big(\tilde{\Pi}_{ \not\!q \gamma_{\mu}} m_{N_2}^{*}({\mu_1}^2-{\mu_2}^2-2 Q^2)+ \tilde{\Pi}_{  \gamma_{\mu}} (m_{N_1}^{*} m_{N_2}^{*}+{\mu_1}^2-2 p_0 \Sigma_{\nu} +\Sigma_{\nu}^2) \big)}{ f_{\rho}^* m_{\rho}^* \lambda_{N_2}^* \lambda_{N_1}^*  \big(m_{N_1}^{*2} m_{N_2}^{*2} + m_{N_1}^{*} m_{N_2}^{*} Q^2 + (-{\mu_2}^2 - Q^2 + 2 p_0 \Sigma_{\nu} - \Sigma_{\nu} ^2) ({\mu_1}^2 - 
		2 p_0 \Sigma_{\nu}  + \Sigma_{\nu} ^2)\big)},
	\label{eq:g1}
\end{eqnarray}
and
\begin{eqnarray}
	&g_2&(\rho_N,Q^2)= \frac{-2 e^{\frac{{\mu_1}^2}{M^2}}e^{\frac{{\mu_2}^2}{M'^2}} (m_{\rho}^{*2} + Q^2)  \tilde{\Pi}_{ \not\!q \gamma_{\mu}} m_{N_2}^{*}({\mu_1}^2-{\mu_2}^2-2 Q^2) (m_{N_1}^{*} m_{N_2}^{*}-{\mu_2}^2-Q^2+2 p_0 \Sigma_{\nu} -\Sigma_{\nu}^2)}{ f_{\rho}^* m_{\rho}^* \lambda_{N_2}^* \lambda_{N_1}^*  ({\mu_1}^2-{\mu_2}^2-2 Q^2) \big(-m_{N_1}^{*2} m_{N_2}^{*2} - m_{N_1}^{*} m_{N_2}^{*} Q^2 + ({\mu_2}^2 + Q^2 -2 p_0 \Sigma_{\nu} + \Sigma_{\nu} ^2) ({\mu_1}^2 - 
		2 p_0 \Sigma_{\nu}  + \Sigma_{\nu} ^2)\big)}	\nonumber \\
	&+	& \frac{-2 e^{\frac{{\mu_1}^2}{M^2}}e^{\frac{{\mu_2}^2}{M'^2}} (m_{\rho}^{*2} + Q^2)\tilde{\Pi}_{  \gamma_{\mu}} \big(2 m_{N_1}^{*2} m_{N_2}^{*2} + m_{N_1}^{*} m_{N_2}^{*} ({\mu_1}^2- {\mu_2}^2)-2({\mu_1}^2  -2 p_0 \Sigma_{\nu} +\Sigma_{\nu}^2) ({\mu_2}^2 +Q^2 -2 p_0 \Sigma_{\nu} +\Sigma_{\nu}^2) \big) }{ f_{\rho}^* m_{\rho}^* \lambda_{N_2}^* \lambda_{N_1}^*  ({\mu_1}^2-{\mu_2}^2-2 Q^2) \big(-m_{N_1}^{*2} m_{N_2}^{*2} - m_{N_1}^{*} m_{N_2}^{*} Q^2 + ({\mu_2}^2 + Q^2 -2 p_0 \Sigma_{\nu} + \Sigma_{\nu} ^2) ({\mu_1}^2 - 
		2 p_0 \Sigma_{\nu}  + \Sigma_{\nu} ^2)\big)},
	\label{eq:g2}
\end{eqnarray}
where $\tilde{\Pi}_i$ denotes the Borel-transformed counterpart of the function $\Pi_i$.
Since the initial and final baryons are identical (both are nucleons), we adopt the symmetric choice for the Borel parameters, $M^2 = M'^2$. Consequently, these parameters are related to the characteristic Borel mass $M_B^2$ via the relation:
\begin{equation}
	M_B^{2} = \frac{M^2 M'^2}{M^2 + M'^2} \implies M^2 = M'^2 = 2M_B^2.
\end{equation}
This symmetric assignment is physically justified, as the double spectral density of the leading-order contributions is predominantly concentrated in the vicinity of the diagonal $s = s'$. Such a configuration ensures stability and minimizes uncertainties, providing a robust framework for the subsequent extraction of the coupling constant.

\section {Numerical Analysis of the Strong Coupling Constants in medium}\label{sec:three}	

In this section, we proceed to the numerical evaluation of the sum rules for the $NN\rho$ strong coupling constant, derived in the preceding section for the nuclear matter environment. To handle the condensates in a dense medium ($\langle {\cal O}\rangle_{\rho_N}$), we utilize the low-density expansion, which relates them to their vacuum expectation values ($\langle {\cal O}\rangle_0$) and the matrix elements between single-nucleon states ($\langle {\cal O}\rangle_N$) as follows:
\begin{equation}
	\langle {\cal O}\rangle_{\rho_N}=\langle 0| {\cal O}|0\rangle+\rho_N\langle N| {\cal O}|N\rangle
	=\langle  {\cal O}\rangle_0+\rho_N\langle {\cal O}\rangle_N.
\end{equation}
Table~\ref{tab:Parameter} summarizes these numerical values utilized for the input parameters in our analysis.

\begin{table}[ht!]
	\centering
	\begin{tabular}{cc}
		\hline \hline
		Input Parameters  &  Values    
		\\
		\hline \hline
		$p_0   $          &  $938.27 \,\text{MeV}$      \\
			$p'_0   $          &  $938.27  \,\text{MeV}$      \\
		$ \rho_N^{sat}   $          &  $(0.11)^3 \,\text{GeV}^3$        \\
		$ \langle q^{\dag}q\rangle_{\rho_N}    $          &  $\frac{3}{2}\rho_{N}$         \\
		$ \langle\bar{q}q\rangle_{0}           $          &  $ (-0.241)^3  \,\text{GeV}^3$          \\
		$ m_{q}      $          &  $0.5(m_{u}+m_{d})$                 \\
		$ \sigma_{N}            $          &  $0.045 ~ \,\text{GeV} $                  \\
		$  \langle\bar{q}q\rangle_{\rho_N}  $          &  $ \langle\bar{q}q\rangle_{0}+\frac{\sigma_{N}}{2m_
			{q}}
		\rho_{N}$                  \\
		$  \langle q^{\dag}g_{s}\sigma
		Gq\rangle_{\rho_N}  $          &  $ -0.33~\,\text{GeV}^2 \rho_{N}$                  \\
		$  \langle q^{\dag}iD_{0}q\rangle_{\rho_N}  $          &  $0.18 ~\,\text{GeV} \rho_{N}$                  \\
		$  \langle\bar{q}iD_{0}q\rangle_{\rho_N}  $          &  $\frac{3}{2} m_q \rho_{N}\simeq0 $                  \\
		$  m_{0}^{2}  $          &  $ 0.8~\,\text{GeV}^2$                  \\
		$   \langle\bar{q}g_{s}\sigma Gq\rangle_{0} $          &  $m_{0}^{2}\langle\bar{q}q\rangle_{0} $                  \\
		$  \langle\bar{q}g_{s}\sigma
		Gq\rangle_{\rho_N}  $          &  $\langle\bar{q}g_{s}\sigma Gq\rangle_{0}+3~\,\text{GeV}^2\rho_{N} $                  \\
		$ \langle  \bar{q}iD_{0}iD_{0}q\rangle_{\rho_N} $          &  $ 0.3~\,\text{GeV}^2\rho_{N}-\frac{1}{8}\langle\bar{q}g_{s}
		\sigma
		Gq\rangle_{\rho_N}$                  \\
		$  \langle
		q^{\dag}iD_{0}iD_{0}q\rangle_{\rho_N}  $          &  $0.031~\,\text{GeV}^2\rho_{N}-\frac{1}{12}\langle
		q^{\dag}g_{s}
		\sigma Gq\rangle_{\rho_N} $                  \\
		$\langle \frac{\alpha_s}{\pi} G^{2}\rangle_{0}$ & $(0.33\pm0.04)^4~\,\text{GeV}^4$\\
		$\langle \frac{\alpha_s}{\pi} G^{2}\rangle_{\rho_N}$ & $\langle \frac{\alpha_s}{\pi} G^{2}\rangle_{0}-0.65~\,\text{GeV} \rho_N$\\
		\hline \hline
	\end{tabular}
	\caption{  Input parameters and their numerical values\cite{Cohen:1994wm,Cohen:1991nk,Jin:1992id}.   }
	\label{tab:Parameter}
\end{table}
In the present analysis, the density dependence of all hadronic parameters entering the sum rules is explicitly taken into account. In particular, the in-medium nucleon mass, nucleon residue, meson mass, and meson decay constant are treated as functions of the nuclear matter density. The in-medium nucleon mass, $m_N^*(\rho_N)$, is determined from the density dependence of the nucleon effective mass presented in Ref.~\cite{Azizi:2014yea}. In particular, the corresponding curve is digitized and fitted with an exponential function, and the resulting parametrization is subsequently employed in the present calculation to incorporate the explicit dependence of the nucleon mass on the nuclear matter density:
\begin{eqnarray}
m_N^*(\rho_N)&=&m_N e^{(-0.459\frac{\rho_N}{\rho_N^{sat}}  )} ,
\end{eqnarray}
 while the in-medium nucleon residue, $\lambda_N^*(\rho_N)$, is incorporated through the density-dependent function given in Ref.~\cite{Mallik:2009jm}
 \begin{eqnarray}
\lambda_N^*(\rho_N)&=&\lambda_N (1-0.20\frac{\rho_N}{\rho_N^{sat}}  ),
 \end{eqnarray}
 where $\rho_N^{sat}$ corresponds to the nuclear matter saturation density.
 The in-medium meson mass is given by \cite{Cohen:1994wm}:
\begin{eqnarray}
m_{\rho}^*(\rho_N)=m_{\rho} (1-0.18\frac{\rho_N}{\rho_N^{sat}}  ),
\end{eqnarray}
  To account for the in-medium modification of the $\rho$ meson decay constant, we adopt the density dependence of the  longitudinal decay constant obtained in the light-front quark model study of Ref.~\cite{Gautam:2026ohc}. Using the numerical results reported in that work, we perform a linear fit to parameterize its dependence on the  density of medium.  The resulting parametrization is given by
  \begin{equation}
  	f_\rho^{  *}(\rho_N)
  	= f_\rho 
  	 (1-0.015 \frac{\rho_N}{\rho_N^{sat}}   ),
  	\label{eq:frho_density}
  \end{equation}
  In the above expressions, $m_N=938.27 \,\text{MeV}$ and $m_{\rho}=770 \,\text{MeV}$ denote the vacuum mass of the nucleon and meson, respectively  \cite{ParticleDataGroup:2022pth}, whereas $\lambda_{N}^2=(0.0012 \pm 0.0006)\,\text{GeV}^6$ \cite{Ioffe:1981kw} and $f_{\rho}=(216 \pm 3) \,\text{MeV}$  \cite{ParticleDataGroup:2024cfk} represent their corresponding vacuum residue and decay constant. Consequently, all the relevant hadronic input parameters are consistently incorporated into the analysis with their explicit dependence on the nuclear matter density. The nucleon  time-like vector self-energy in the dense nuclear medium is given by   $\Sigma_0=(420\pm65)\,\text{MeV}$  \cite{Azizi:2014yea}. 
  Note that in the rest frame of the nuclear medium, where the four-velocity is defined as  $u^\mu = (1, \mathbf{0})$, the spatial parts of the vector self-energy vanish, and its time component reduces to $\Sigma_{\nu} \equiv \Sigma_0 = (420 \pm 65)~\text{MeV}$. 	The light quark masses are taken to be $m_u =  (2.16\pm0.04)\,\text{MeV}$ and $m_d = (4.7\pm0.04)\,\text{MeV}$, respectively \cite{ParticleDataGroup:2022pth}.

In addition to these input parameters, the sum rule expressions involve three auxiliary parameters: the continuum threshold $s_0$, the Borel mass parameter $M_B^2$, and the mixing parameter $\beta$ introduced through the nucleon interpolating current. These parameters are constrained by enforcing the standard requirements of the QCD sum rule framework, namely, stability against variations of the auxiliary parameters, pole dominance, and convergence of the operator product expansion. 
Guided by these physical criteria, the continuum threshold is restricted to the interval:
\begin{eqnarray}
	1.50\,\,\text{GeV}^{2}&\leq& s_{0}\leq 2.50\,\,\text{GeV}^{2}.
\end{eqnarray}
Furthermore, the upper boundaries of the Borel parameters are established by requiring a sufficient pole contribution within the designated working windows of the continuum thresholds. To this end, we examine the following ratio constructed from the continuum-subtracted and Borel-transformed invariant amplitude $\tilde{\Pi}_i(s_0,M_B^2,\beta)$ evaluated on the QCD side:
\begin{eqnarray}
	PC=\frac{\tilde{\Pi}_i(s_0,M_B^2,\beta)}{\tilde{\Pi}_i(\infty,M_B^2,\beta)}.
\end{eqnarray}
Here, $PC$ represents the pole contribution, while $i$ indexes the chosen Lorentz structures. The upper bound of the Borel window is established by ensuring that this ratio remains at least $50\%$.

Conversely, the lower limit of $M_B^2$ is determined by ensuring the dominance of the perturbative series over non-perturbative contributions, alongside the requirement that higher-dimensional operator terms remain suppressed, thereby guaranteeing the convergence of the OPE series. These constraints restrict the Borel mass squared to the working range of
\begin{eqnarray}
0.50~\text{GeV}^2 \leq M_B^2 \leq 0.90~\text{GeV}^2.
\end{eqnarray}

\begin{figure}[h!]
	\begin{center}
		\includegraphics[totalheight=6.5cm,width=8.5cm]{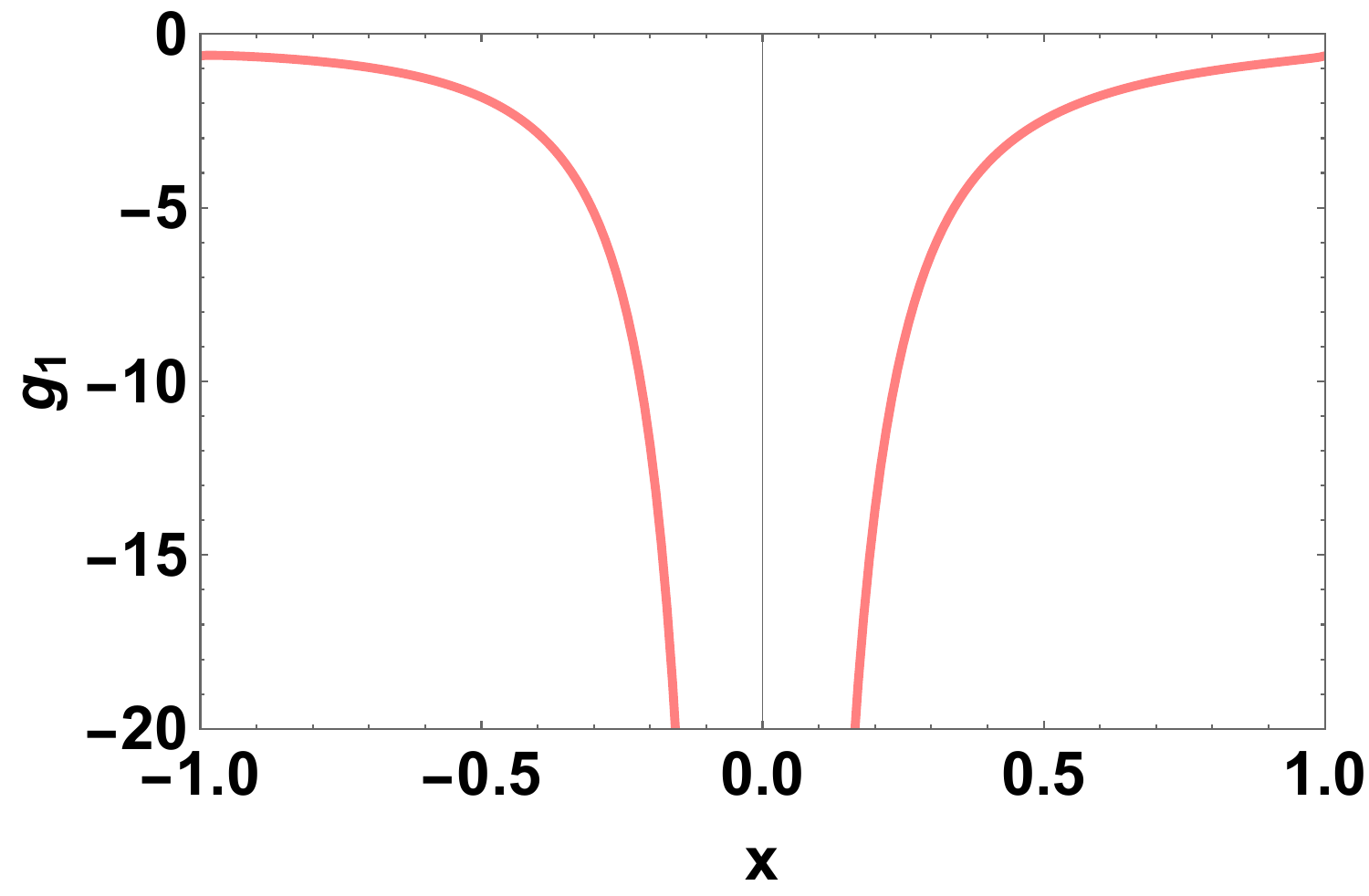}
		\includegraphics[totalheight=6.5cm,width=8.5cm]{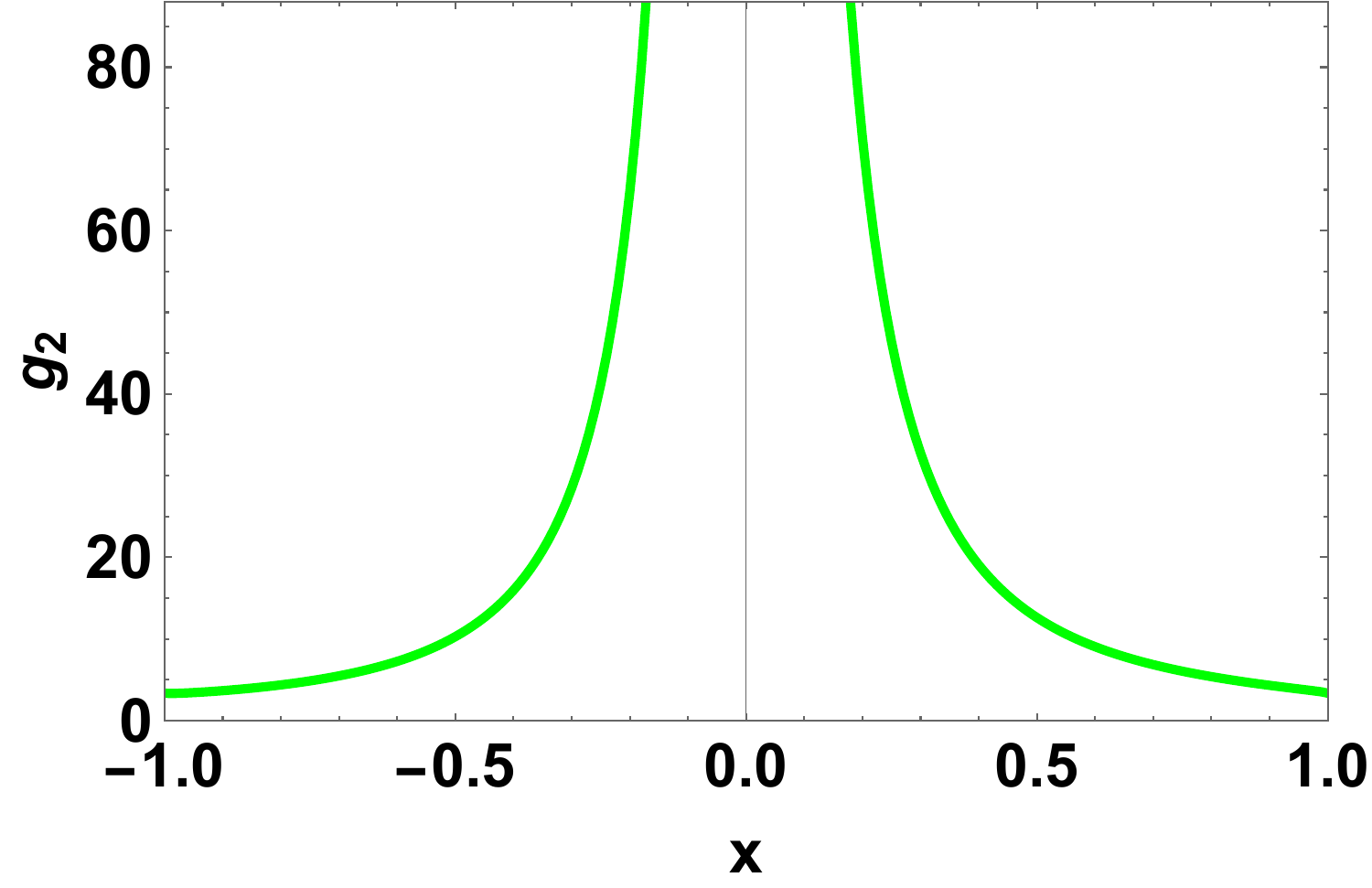}
	\end{center}
\caption{Left: Dependence of the strong coupling constant $g_1$ on $x = \cos\theta$ at fixed values of the continuum threshold and Borel parameter. 
	Right: Dependence of the strong coupling constant $g_2$ on $x = \cos\theta$ under the same conditions.}
	\label{x}
\end{figure}

\begin{figure}[h!]
	\begin{center}
		\includegraphics[totalheight=6.5cm,width=8.5cm]{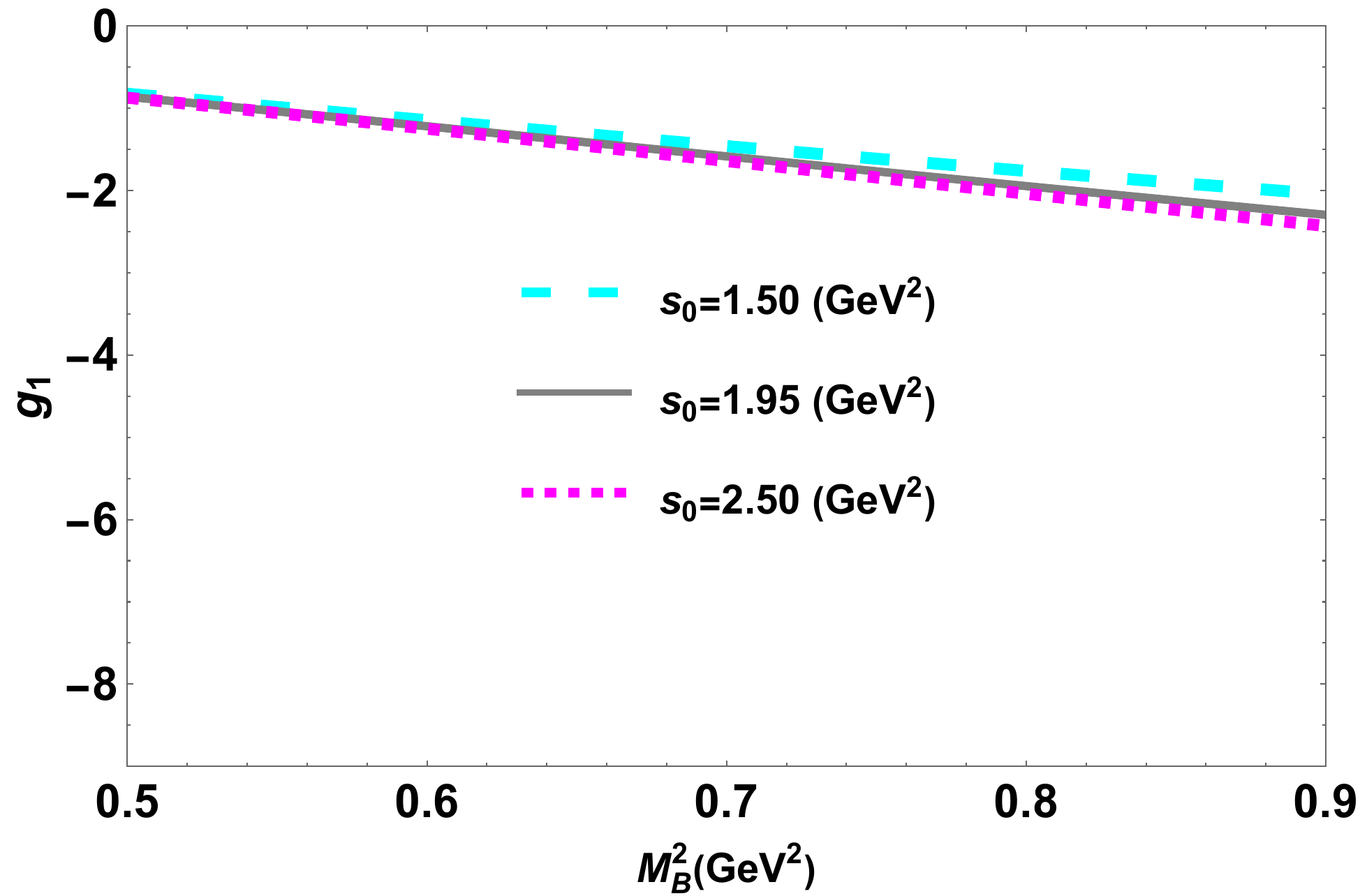}
		\includegraphics[totalheight=6.5cm,width=8.5cm]{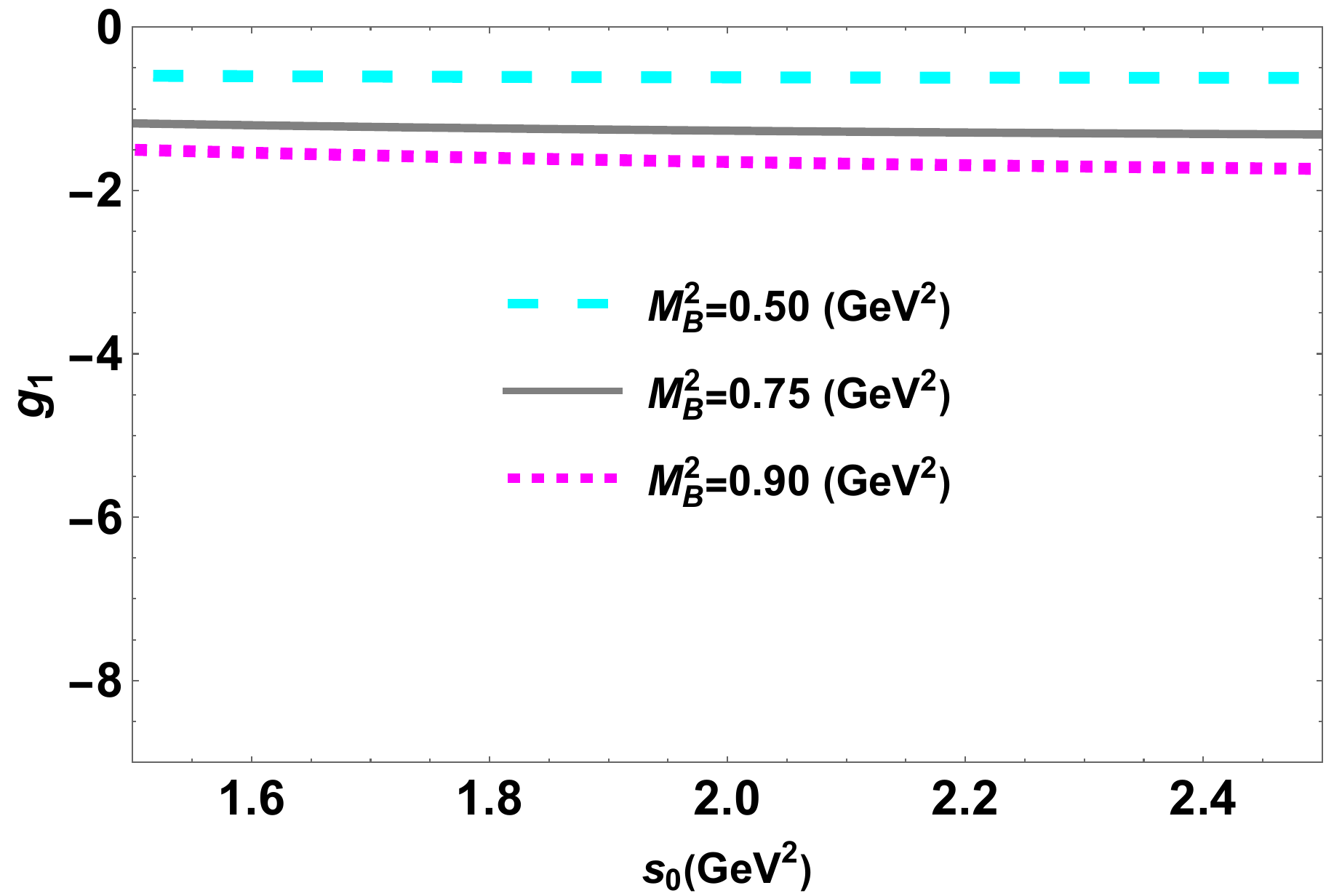}
	\end{center}
	\caption{Left: Behavior  of the extracted $g_{1}$ as a function of $M_B^2$ at
		various values of continuum thresholds, at Ioffe point  $\cos\theta =-0.71$ and $Q^2=-m_{\rho}^{*2} $. \textbf{Right:} Behavior  of the extracted  $g_{1}$ as
		a function of $s_0$ at various values of Borel mass parameter, at Ioffe point 
		$\cos\theta =-0.71$ and $Q^2=-m_{\rho}^{*2}$.}
	\label{g1}
\end{figure}

\begin{figure}[h!]
	\begin{center}
		\includegraphics[totalheight=6.5cm,width=8.5cm]{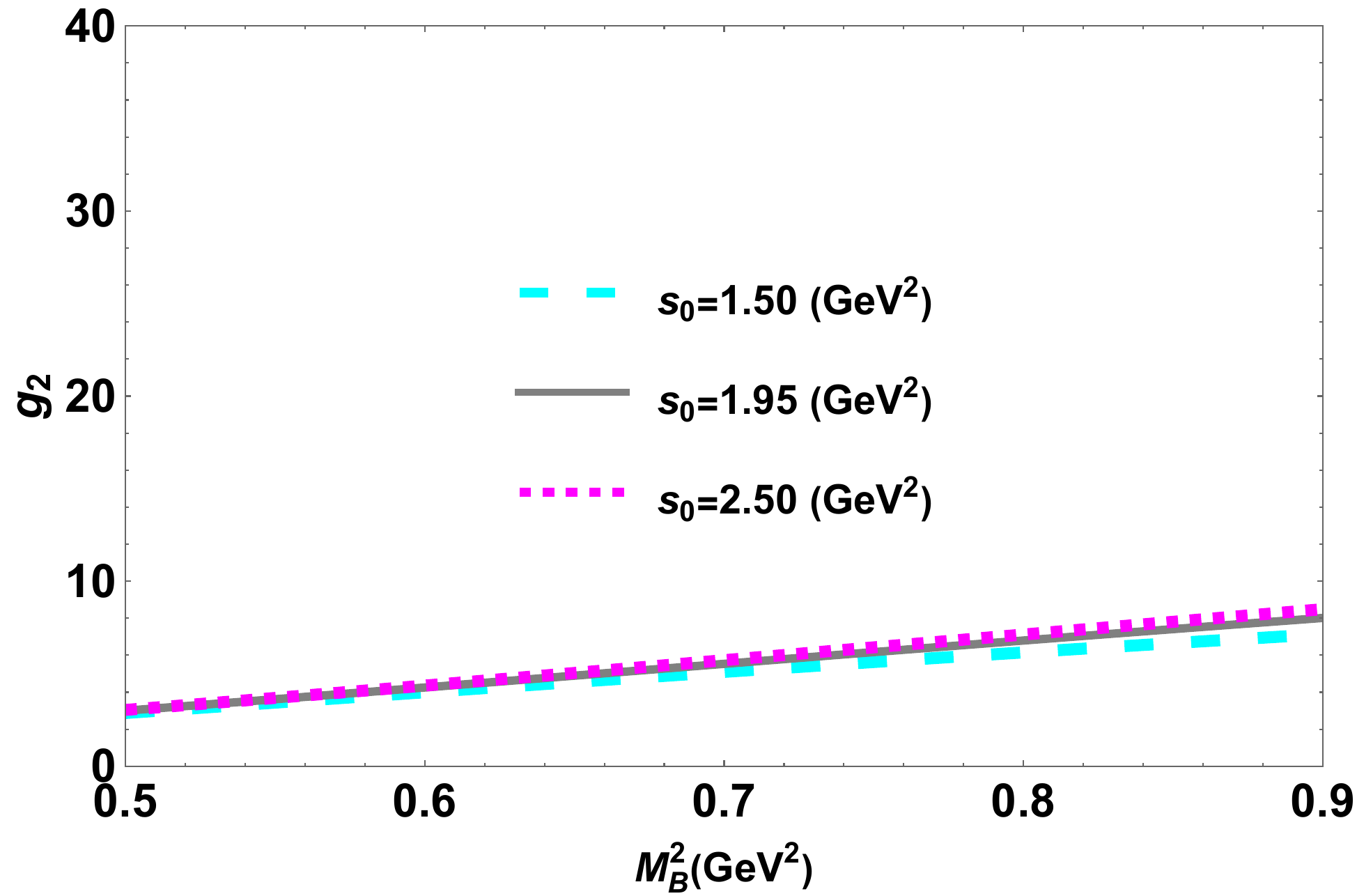}
		\includegraphics[totalheight=6.5cm,width=8.5cm]{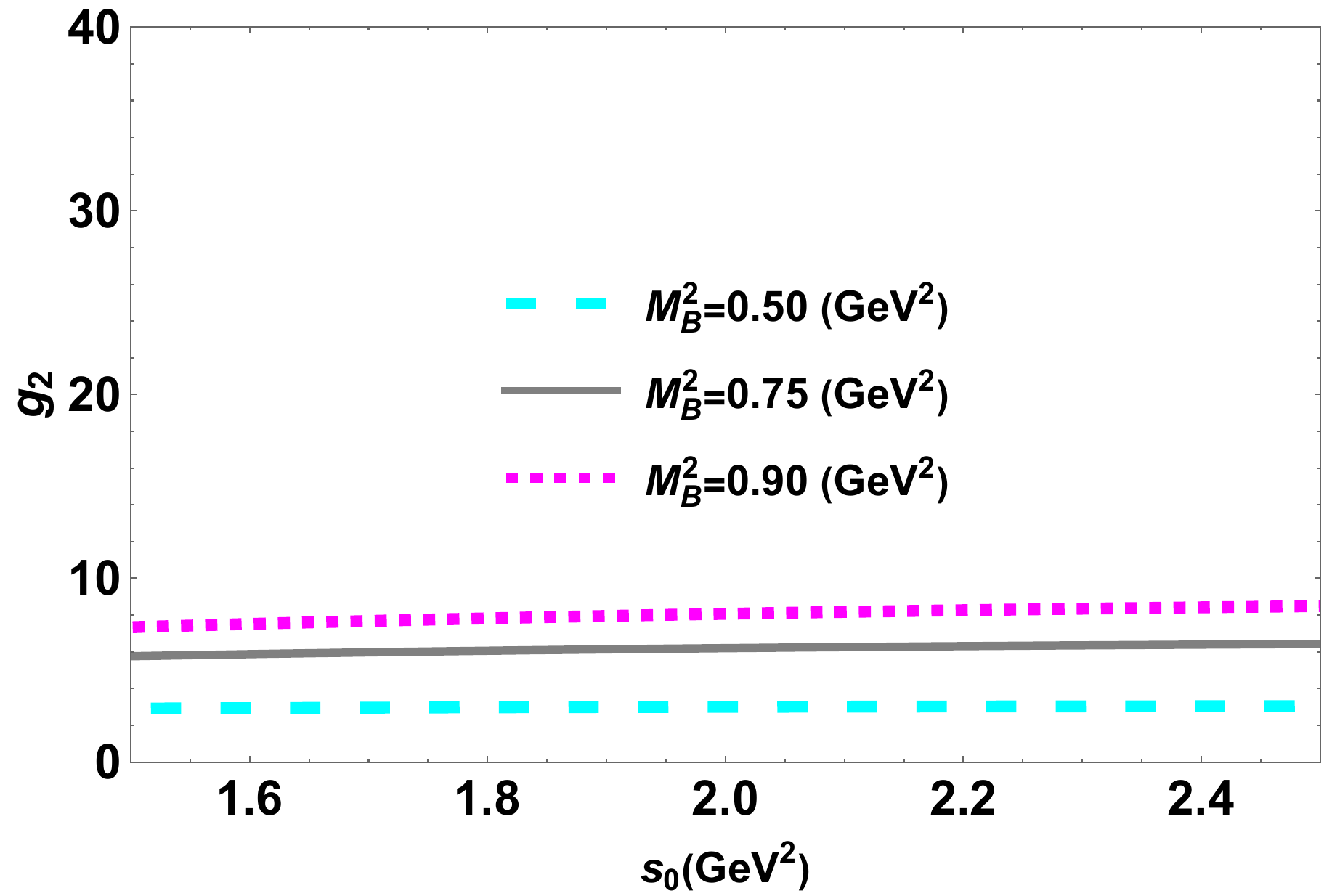}
	\end{center}
	\caption{Left: Behavior  of the extracted $g_{2}$ as a function of $M_B^2$ at
		various values of continuum thresholds, at Ioffe point  $\cos\theta =-0.71$ and $Q^2=-m_{\rho}^{*2} $. \textbf{Right:} Behavior  of the extracted  $g_{2}$ as
		a function of $s_0$ at various values of Borel mass parameter, at Ioffe point 
		$\cos\theta =-0.71$ and $Q^2= -m_{\rho}^{*2} $.}
	\label{g2}
\end{figure}

Finally, the optimal working interval for the mixing parameter $\beta$ is identified by analyzing the parametric dependence of the results on $\theta$, where $\beta = \tan\theta$. We examine the behavior of $g_1$ and $g_2$ as functions of $\cos\theta$ at specific, fixed values of the continuum threshold and Borel parameter, as illustrated in Fig.~\ref{x}. Our selection criteria rely on identifying stability regions-intervals where the results exhibit minimal sensitivity to variations in $\cos\theta$.
As illustrated in these figures, both $g_1$ and $g_2$ exhibit remarkable stability, remaining essentially independent of the parameter $x$ within the intervals $-1\leq x \leq -0.5$ and $0.5\leq x \leq 1$. Furthermore, the dependence of these coupling constants on the auxiliary sum rule parameters is depicted in Figs.~\ref{g1} and \ref{g2}, evaluated at the kinematic point $Q^2=-m_{\rho}^{*2}$.

Notably, the Ioffe current, corresponding to $\cos\theta=-0.71$, lies within the adopted working intervals. The numerical analysis is performed over the considered range
of nuclear matter densities, while the stability analysis with
respect to the auxiliary sum rule parameters is carried out
at the saturation density, $\rho_N^{sat}$. An inspection of the figures and the corresponding numerical data confirms the adequate stability of the extracted results against variations of the continuum threshold and the Borel parameter throughout their respective working windows.

Armed with the specified input parameters and the established intervals for the auxiliary variables, we proceed to compute the strong coupling constants for the transition under investigation. The $Q^2$-dependence of the strong coupling form factors is subsequently parametrized through the following fit expression:

\begin{eqnarray}
	g_i(Q^2)&=& g_{0}e^{c_1\frac{Q^2}{m_{N}^{*2}}+c_2(\frac{Q^2}{m_{N}^{*2}})^2}.
\end{eqnarray}
In this expression, $g_0$, $c_1$, and $c_2$ represent the fit parameters, with their numerical values summarized in Table~\ref{tab:FitParam} at the saturation density, $\rho_N^{sat}$. 
\begin{table}[tbp]
	\begin{tabular}{|c|c|c|c|}
		\hline\hline Coupling Constant  &$ g_0$ & $c_1$ & $c_2$ \\ \hline\hline
		$g_1$&$-1.315 \pm 0.211$ &$0.044$ & $-0.002$ \\
		\hline
		$g_2$&$7.315\pm 0.631$ & $-0.007$ & $0.0006$ \\
		\hline\hline
	\end{tabular}%
	\caption{The fit   parameters  for the
		coupling constants, 	$g_i(Q^2)$.} \label{tab:FitParam}
\end{table}
We subsequently utilize this function to extrapolate the strong coupling constants to the kinematic point $Q^2=-m_{\rho}^{*2}$, yielding:
\begin{equation}
	g_1 = -1.379 \pm 0.231, \qquad g_2 = 7.377 \pm 0.771.
\end{equation}
The quoted uncertainties account for the aggregate errors originating from the input QCD parameters, the selection of auxiliary sum rule variables, and the fitting procedure itself.

\begin{figure}[h!]
	\begin{center}
		\includegraphics[totalheight=6.5cm,width=8.5cm]{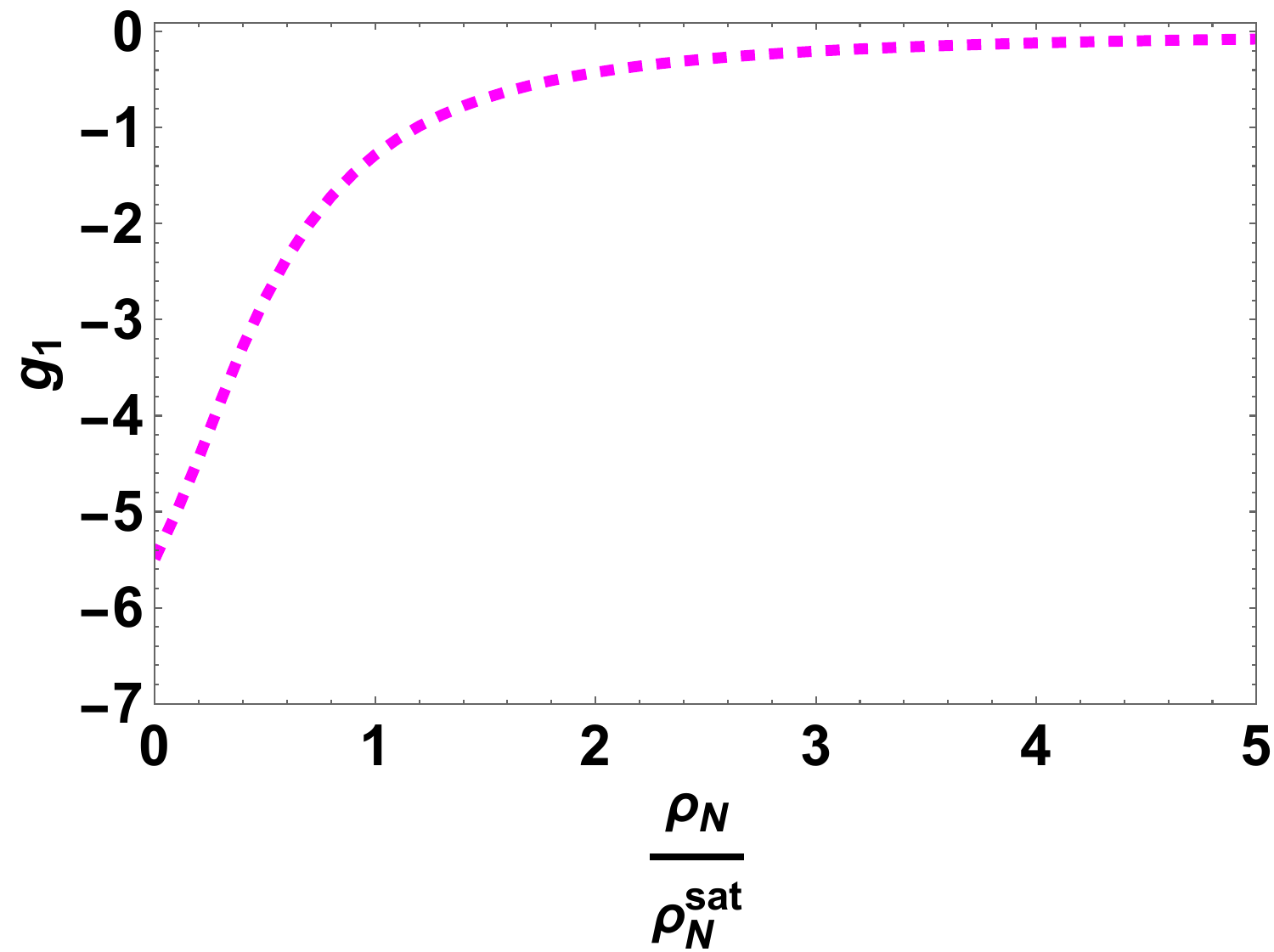}
		\includegraphics[totalheight=6.5cm,width=8.5cm]{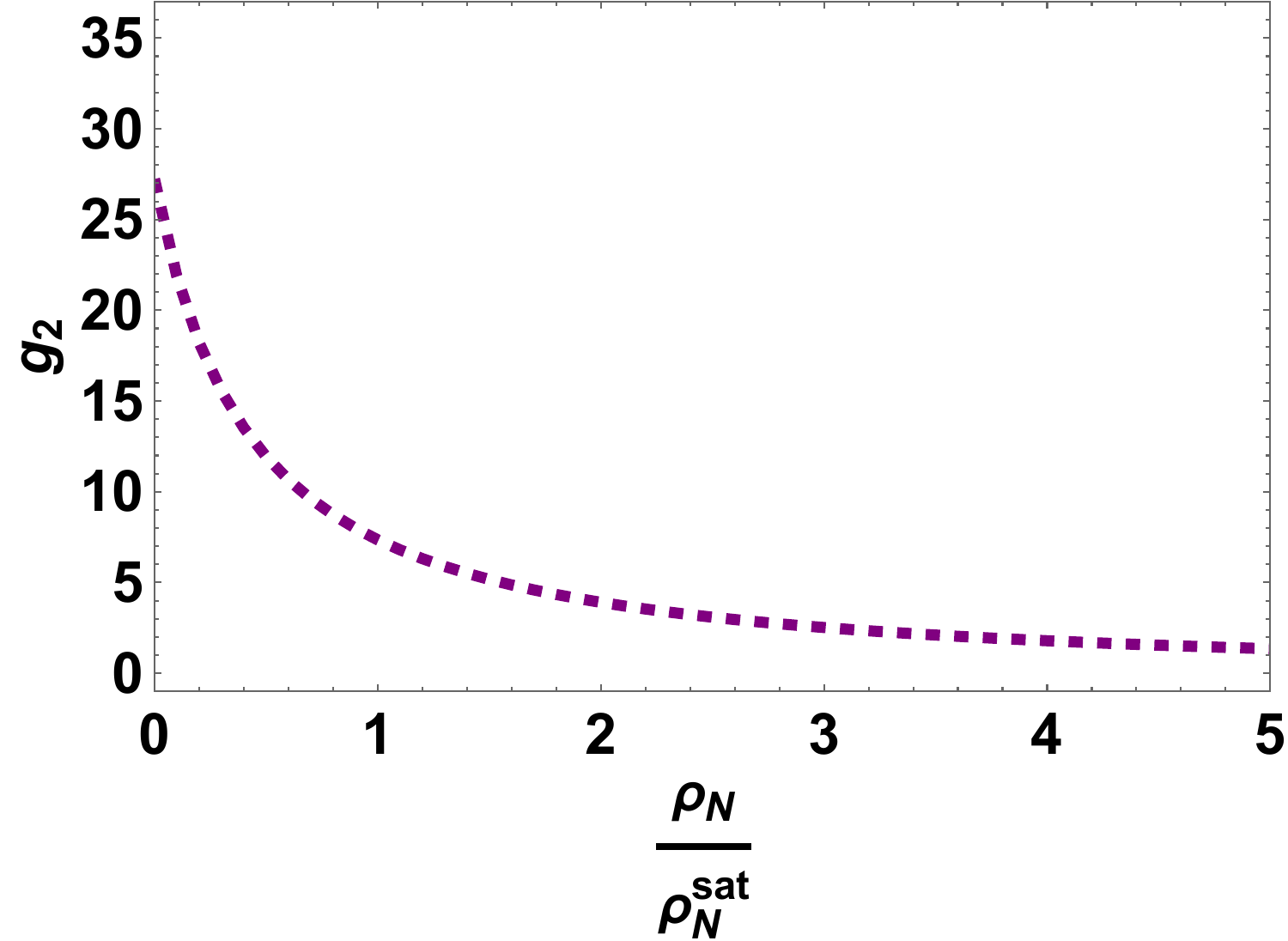}
	\end{center}
	\caption{Left: Dependence of the strong coupling constant $g_1$ on $\frac{\rho_N}{\rho_N^{sat}} $ at fixed values of the continuum threshold and Borel parameter, at Ioffe point 
		$\cos\theta =-0.71$ and $Q^2= -m_{\rho}^{*2} $. 
		Right: Dependence of the strong coupling constant $g_2$ on $\frac{\rho_N}{\rho_N^{sat}} $ under the same conditions.}
	\label{1}
\end{figure}

Finally, it is instructive to investigate the density dependence of the strong coupling constants $g_1$ and $g_2$, associated with the $NN\rho$ vertex at $Q^2= -m_{\rho}^{*2} $. Our numerical results indicate that the density dependence of these coupling constants can be described satisfactorily by the following fitting functions:

\begin{eqnarray}\label{fitfunc}
	g_{i}(\rho_{N})=\frac{g'_0}{1-a \frac{\rho_N}{\rho_N^{sat}}+b (\frac{\rho_N}{\rho_N^{sat}})^2},
\end{eqnarray}
where  $g'_0$, $a$, and $b$ are the fit parameters, with their numerical values summarized in Table~\ref{tab:FitParam2}.  
\begin{table}[tbp]
	\begin{tabular}{|c|c|c|c|}
		\hline\hline Coupling Constant  &$ g'_0$ & $a$ & $b$ \\ \hline\hline
		$g_1$&$-5.417 \pm 0.559$ &$-0.288$ & $2.645$ \\
		\hline
		$g_2$&$27.234\pm 2.638$ & $-2.419$ & $0.284$ \\
		\hline\hline
	\end{tabular}%
	\caption{The fit   parameters  for the
		coupling constants,	$g_{i}(\rho_{N})$.} \label{tab:FitParam2}
\end{table}
The corresponding results for the strong coupling constants as functions of the $ \frac{\rho_N}{\rho_N^{sat}} $ are presented in Fig.~\ref{1}.

In the vacuum limit, corresponding to $\rho_{N}=0$, our results yield $g_1 = -5.417 \pm 0.559$ and $g_2 = 27.234 \pm 2.638$. These values are in good agreement with the results reported in Ref.~\cite{Aliev:2009ei}, which found $g_1 =-5.9\pm 1.3  $ and $g_1 + g_2 = 22.7\pm 1.3$, respectively, thereby providing a consistency check on our calculations.
 The density-dependent coupling constants obtained in this work may serve as useful inputs for further investigations of hadronic interactions in nuclear matter and other dense hadronic environments. In particular, they may be relevant to studies of neutron-star matter, where modifications of hadronic properties induced by finite density can influence the description of the underlying strong interactions. Our results can therefore provide a useful reference for future theoretical analyses of dense nuclear matter and neutron-star structure.

\section {Conclusion}\label{sec:four}

In this study, we investigated the in-medium strong coupling constants governing the $NN\rho$ interaction within the framework of QCD sum rules. While previous sum rule applications in nuclear matter have predominantly focused on the static properties of isolated hadrons, such as in-medium masses and decay constants, this work provides a systematic QCD-based evaluation of a two-body hadronic interaction vertex at finite nuclear density.

To this end, we formulated a three-point in-medium correlation function composed of the interpolating currents of the nucleon and the $\rho$ meson. On the phenomenological side, the interaction was parameterized in terms of coupling form factors, $g_1$ and $g_2$, alongside the density-dependent modifications of the hadron masses, residues and decay constants. On the theoretical side, the short-distance operator product expansion was evaluated by retaining both perturbative corrections and the relevant non-perturbative contributions  in the nuclear medium, encompassing quark, gluon, and mixed condensates under the linear density approximation.
By matching the analytical structures of the hadronic  with the OPE representation and applying a double Borel transformation, the sum rules for $g_1$ and $g_2$ were extracted. 

The  $Q^2$-dependence of the resulting form factors was parameterized using exponential fit functions, facilitating an extrapolation to the physical on-shell point $Q^2 = -m_{\rho}^{*2}$ at nuclear matter saturation density $\rho_N^{\text{sat}}$. At the on-shell point, we obtained the strong coupling constants, with uncertainties that account for the estimated contributions from the QCD input parameters, the in-medium condensates, and the variations of the auxiliary sum rule parameters. 
To conclude, we have calculated the $NN\rho$ vertex coupling constants $g_1$ and $g_2$ as functions of nuclear matter density via QCD sum rules. Our vacuum results ($g_1 = -5.417 \pm 0.559$ and $g_2 = 27.234 \pm 2.638$) show consistency with prior studies  ($g_1 = -5.9 \pm 1.3$ and $g_1 + g_2 = 22.7 \pm 1.3$) \cite{Aliev:2009ei}. The density evolution of $g_1$ and $g_2$ was successfully mapped out using dedicated fitting functions with parameters listed in Table~\ref{tab:FitParam2}. Our findings provide a QCD sum rule-based assessment of the in-medium  $NN\rho$ interaction and may serve as useful microscopic input for phenomenological, experimental and observational studies of dense hadronic matter, including the neutron-star matter.

\section*{Funding Declaration} 
This work is based upon research funded by  Iran National Science Foundation (INSF)  under project No. 40408515.


\end{document}